\documentclass[%
 reprint,
superscriptaddress,
 amsmath,amssymb,
 aps,prl,
floatfix,
]{revtex4-2}

\usepackage{graphicx}% Include figure files
\usepackage{dcolumn}% Align table columns on decimal point
\usepackage{bm}% bold math
\usepackage{braket}
\usepackage{hyperref}% add hypertext capabilities
\hypersetup{pdfpagemode=FullScreen,
colorlinks=true, citecolor = blue, linkcolor=blue, filecolor=blue}
\usepackage[mathlines]{lineno}% Enable numbering of text and display math
\usepackage{amsmath}
\usepackage[T1]{fontenc}
\begin{document}
%\linenumbers
\preprint{APS/123-QED}

\title{Enhanced quantum sensing with room-temperature solid-state masers}% Force line breaks with \\

\author{Hao Wu}
\altaffiliation[]{These authors contributed equally.}
\affiliation{% 
Center for Quantum Technology Research and Key Laboratory of Advanced Optoelectronic Quantum Architecture and Measurements (MOE), School of Physics, Beijing Institute of Technology, Beijing 100081, China
}%
\affiliation{% 
Beijing Academy of Quantum Information Sciences, Beijing 100193, China
}%

\author{Shuo Yang}%
\altaffiliation[]{These authors contributed equally.}
\affiliation{% 
Center for Quantum Technology Research and Key Laboratory of Advanced Optoelectronic Quantum Architecture and Measurements (MOE), School of Physics, Beijing Institute of Technology, Beijing 100081, China
}%
\affiliation{% 
Beijing Academy of Quantum Information Sciences, Beijing 100193, China
}%

\author{Mark Oxborrow}
\affiliation{%
Department of Materials, Imperial College London, South Kensington SW7 2AZ, London, United Kingdom}%

\author{Qing Zhao}
\affiliation{% 
Center for Quantum Technology Research and Key Laboratory of Advanced Optoelectronic Quantum Architecture and Measurements (MOE), School of Physics, Beijing Institute of Technology, Beijing 100081, China
}%
\affiliation{% 
Beijing Academy of Quantum Information Sciences, Beijing 100193, China
}%

\author{Bo Zhang}
\email{bozhang_quantum@bit.edu.cn}
\affiliation{% 
Center for Quantum Technology Research and Key Laboratory of Advanced Optoelectronic Quantum Architecture and Measurements (MOE), School of Physics, Beijing Institute of Technology, Beijing 100081, China
}%
\affiliation{% 
Beijing Academy of Quantum Information Sciences, Beijing 100193, China
}%

\author{Jiangfeng Du}
\affiliation{%
CAS Key Laboratory of Microscale Magnetic Resonance and School of Physical Sciences, University of Science and Technology of China, Hefei 230026, China
}%
\affiliation{%
CAS Center for Excellence in Quantum Information and Quantum Physics, University of Science and Technology of China, Hefei 230026, China
}%

\date{\today}% It is always \today, today,

\begin{abstract}
Quantum sensing with solid-state systems finds broad applications in diverse areas ranging from material and biomedical sciences to fundamental physics. Several solid-state spin sensors \cite{barry2020sensitivity,christle2015isolated,rugar2004single} have been developed, facilitating the ultra-sensitive detection of physical quantities such as magnetic and electric fields\cite{barry2020sensitivity,clevenson2015broadband,jensen2014cavity,dolde2011electric,li2020nanoscale} and temperature\cite{kucsko2013nanometre,liu2021ultra}. Exploiting collective behaviour of non-interacting spins holds the promise of pushing the detection limit to even lower levels\cite{clevenson2015broadband}, while to date, those levels are scarcely reached due to the 
broadened linewidth and inefficient readout of solid-state spin ensembles\cite{barry2020sensitivity}. Here, we experimentally demonstrate that such drawbacks can be overcome by newly reborn maser technology\cite{arroo2021perspective} at room temperature in the solid state. Owing to maser action, we observe a 4-fold reduction in the inhomogeneously broadened linewidth of a molecular spin ensemble, which is narrower than the same measured from single spins at cryogenic temperatures. The maser-based readout applied to magnetometry showcases a signal-to-noise ratio (SNR) of 30 dB for single shots. This technique would be a significant addition to the toolbox for boosting the sensitivity of solid-state ensemble spin sensors.

\end{abstract}

\maketitle

%\tableofcontents

%\section{\label{sec:level1}Introduction}

Quantum sensing exploits high sensitivity of quantum systems to external disturbances for measurements of physical quantities. Most recently, solid-state spin systems\cite{wolfowicz2021quantum}, without the need for heating, vacuum or cryogenic conditions, have emerged as an increasingly favorable platform for ultra-sensitive quantum sensing, with advantages over other systems in terms of the robustness\cite{liu2019coherent}, biocompatibility\cite{shi2018single} and spatial resolution\cite{taylor2008high}. 

An intriguing feature of quantum sensors is that their ultimate sensitivity, predominantly determined by quantum-projection noise, can be further improved by increasing the number of non-interacting spins\cite{taylor2008high}. This approach (i.e. employing spin ensembles) is nowadays widely used in solid-state spin sensors for sensitivity enhancement\cite{barry2020sensitivity}. However, there are two obstacles existing towards realizing solid-state ensemble spin sensors with a sensitivity approaching the quantum-projection-noise limit. First, increasing spin density in solids unavoidably introduces complexities and variations in the local environment of individual spins, manifesting as broadened inhomogeneous linewidth of the ensemble.   
%Although certain pulsed techniques such as dynamical decoupling can greatly extend the spin coherence time (ref.), they are mostly applied for fast varying quantities like ac magnetic field.
Solid-state ensemble spin sensors exploiting slope detection will thus struggle to sense slowly varying physical quantities, such as (near-)dc fields and bio-temperature variations\cite{barry2020sensitivity,kucsko2013nanometre}. Although several pulsed techniques, e.g. Ramsey and pulsed optically detected magnetic resonance (pulsed-ODMR)\cite{dreau2011avoiding,barry2020sensitivity}, have been applied to 
eliminate power broadening of the electron spin resonance (ESR), solid-state ensemble spin sensors are also plagued by the lack of efficient readout schemes, which is the second obstacle.  Optical readout based on the detection of spin-state-dependent photoluminescence (PL) is, to date, the most common readout technique, but suffers from low readout fidelities arising from non-negligible photon shot noise. Albeit for spin ensembles, the collected PL signals are magnified , the optical readout fidelity is not necessarily improved since the measurement contrast is degraded by the elevated background luminescence from the spins or impurities not contributing to sensing\cite{rondin2014magnetometry}. Overall, inefficient readout of solid-state ensembles results in the sensors' sensitivities being limited by photon shot noise about two orders of magnitude worse than the quantum-projection-noise limit\cite{barry2020sensitivity}. 
%does not allow single-shot determination of the spin state to the spin-projection limit. Previous studies have shown great improvement of the readout fidelity, but most focus on single spins, limiting the application for high-sensitivity measurements. 

\begin{figure*}[htbp!]
\includegraphics{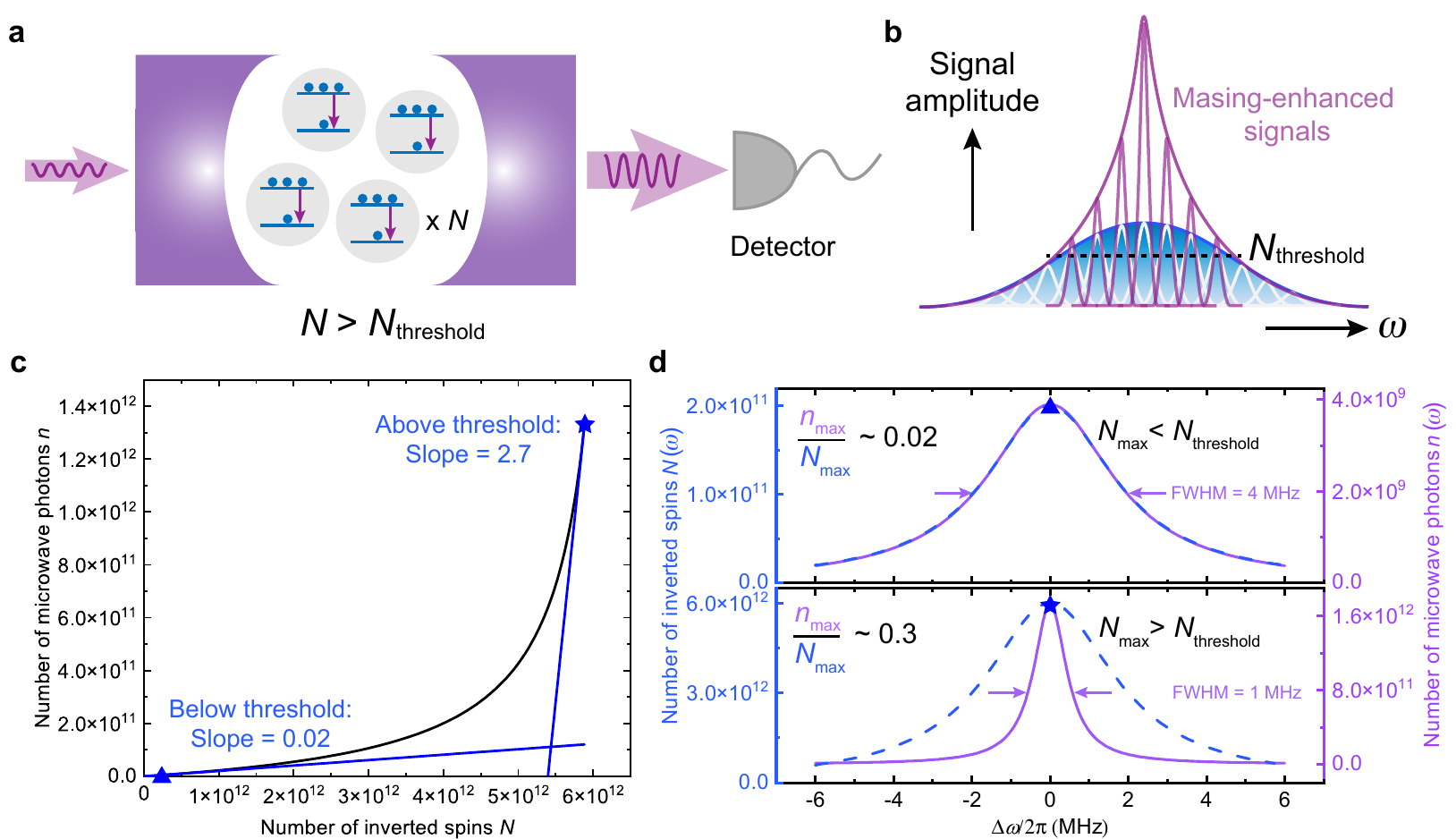}% Here is how to import EPS art
\caption{\label{fig: Concept_Figure_211117}\textbf{Concept and simulation results of the masing-enhanced quantum sensing.} \textbf{a,} General mechanism of masing. Stimulated emission of radiation (downward arrows) occurs when spins with population inversion (blue circles) in a resonator (pink enclosure) interact coherently with injected microwave photons (left-hand-side arrow). The amount of detected microwave photons (right-hand-side arrow) is significantly increased, so-called masing, if the number of the inverted spins $N$ is greater than that required for achieving the masing threshold ($N_\textrm{threshold}$). \textbf{b,} Schematic illustration of the masing effect on ensemble spin resonance. An ensemble spin resonance spectrum (blue curve) constitutes numerous spin packets (white) where inverted spins are resonant at different frequencies. The signal associated with specific spin packets where the amount of inverted spins $N$ is greater than $N_\textrm{threshold}$ can be enhanced by maser action (purple packets) and extracted from the microwave resonator. The observed linewidth is narrowed correspondingly (purple envelope). \textbf{c,} Simulation of the threshold behaviour. The number of microwave photons $n$ in a resonator increases non-linearly with the number of inverted spins $N$. The blue triangle and star are labelled in the below- and above-threshold regimes with the associated tangent lines and slopes where $N = 2\times10^{11}$ and $6\times10^{12}$, respectively. \textbf{d,} Simulation of the masing effect on measured ensemble spin resonance. The distribution of the number of inverted spins across the resonance $N(\omega)$ is set to be Lorentzian (blue dashed lines). By varying the amplitude of the distribution $N_\textrm{max}$ (blue triangle and star from \textbf{c}), the measured spin ensemble resonance in the below- (top panel) and above-threshold (bottom panel) regimes is reflected by distinct distributions of the detected number of microwave photons $n(\omega)$ (purple solid lines) in terms of the amplitude and linewidth. }
\end{figure*}

In this work, we incorporate newly developed room-temperature maser technology\cite{oxborrow2012room,breeze2018continuous,jin2015proposal,jiang2021floquet} into quantum sensing and demonstrate  that the threshold behaviour and microwave amplification of maser action can be exploited in solid-state ensemble spin sensors for simultaneously addressing both issues hereinbefore mentioned. The concept of masing-enhanced quantum sensing is schematically illustrated in Fig.~\ref{fig: Concept_Figure_211117}a and \ref{fig: Concept_Figure_211117}b. We consider a solid-state spin ensemble that comprises $N$ two-level spin systems with microwave transitions and possessing population inversion, so-called inverted spins, which can be achieved by microwave pumping\cite{angerer2018superradiant} or optical hyperpolarization protocols\cite{oxborrow2012room,breeze2018continuous}. Masing occurs when the ensemble is placed in a microwave resonator and the stimulated emission is induced by the injected microwave photons (acting as probe signals) on resonance with the inverted spins, yielding an increased number of microwave photons in the resonator, $n$. When $N$ is sufficiently large and surpasses a threshold value $N_\textrm{threshold}$, the stimulated emission will be dominant over loss mechanisms in the resonator resulting in a surge of detected microwave photons, thus significantly enhancing the readout signals (see Fig.~\ref{fig: Concept_Figure_211117}a). For the solid-state spin ensemble with inhomogeneous broadening, its magnetic resonance lineshape can be regarded as a convolution of spin packets with different resonance frequencies and amplitudes, as shown in Fig.~\ref{fig: Concept_Figure_211117}b. The amplitudes are associated with the number of the inverted spins detected via magnetic resonance. Under the condition that certain spin packets near the line center fulfill the requirements of masing, the detected amplitudes are dramatically enhanced resulting in overall narrowing of the resonance lineshape, which can benefit the slope-detection-based quantum sensing\cite{degen2017quantum}.

\begin{figure*}[htbp!]
\includegraphics{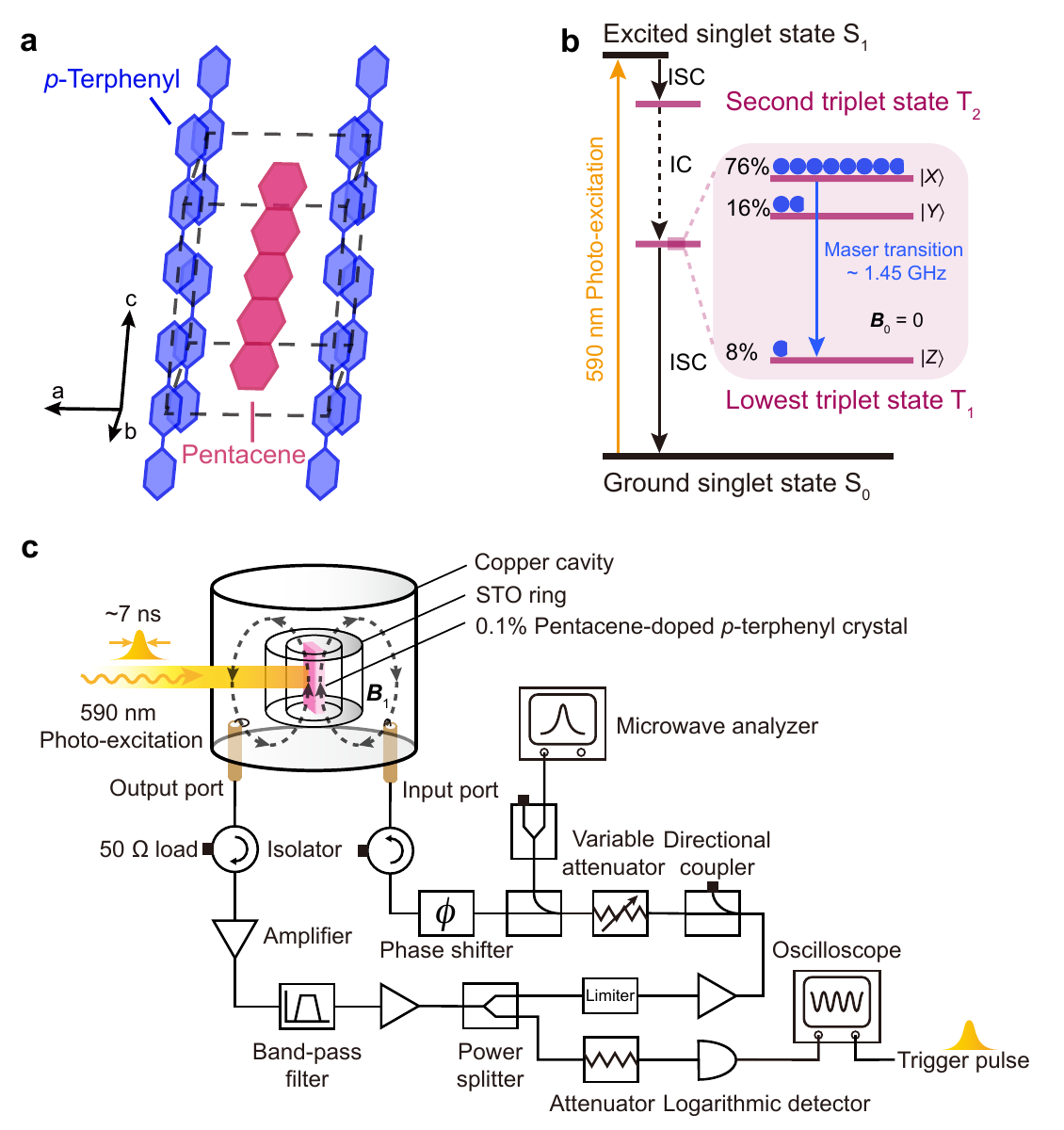}% Here is how to import EPS art
\caption{\label{fig: Spin_system_setup}\textbf{Spin system and experimental setup.} \textbf{a,} Crystal structure of pentacene-doped \textit{p}-terphenyl. A pentacene molecule is doped in \textit{p}-terphenyl's monoclinic unit cell in which a \textit{p}-terphenyl molecule is substituted. \textbf{b,} Simplified Jablonski diagram depicting the origin of pentacene's triplet spins. 590-nm photo-excitation promotes pentacene's singlet spins from the ground state S$_0$ to the first excited state S$_1$. Undergoing intersystem crossing (ISC), the spins are transferred to the second triplet state (T$_2$) and then quickly decay to the lowest triplet state (T$_1$) via internal conversion (IC). The spins in T$_1$ are relatively long-lived before non-radiatively decaying to S$_0$ by ISC. At zero applied magnetic field (\textbf{\textit{B}$_\textrm{0}$} $=0$), T$_1$ is non-degenerate where the three spin sublevels $\ket{X}$, $\ket{Y}$ and $\ket{Z}$ possess population inversions. The distribution of the population in each sublevel is labelled and also indicated by blue circles. Masing with a frequency of $\sim$1.45 GHz can occur between the $\ket{X}$ and $\ket{Z}$ sublevels. \textbf{c,} Experimental setup for lineshape measurements and sensing an external magnetic field. A 0.1\% pentacene-doped \textit{p}-terphenyl crystal (pink) is placed inside a strontium titanate (STO)-based microwave resonator where the TE$_{01\delta}$ mode associated ac magnetic field (\textbf{\textit{B}$_\textrm{1}$}) is depicted. The resonator connected to a feedback loop (all components labelled) is configured into a regenerative microwave oscillator for the measurements. 7-ns laser pulses with a wavelength of 590 nm are employed to excite the crystal and to trigger the oscilloscope. }
\end{figure*}

\begin{figure*}[htbp!]
\includegraphics{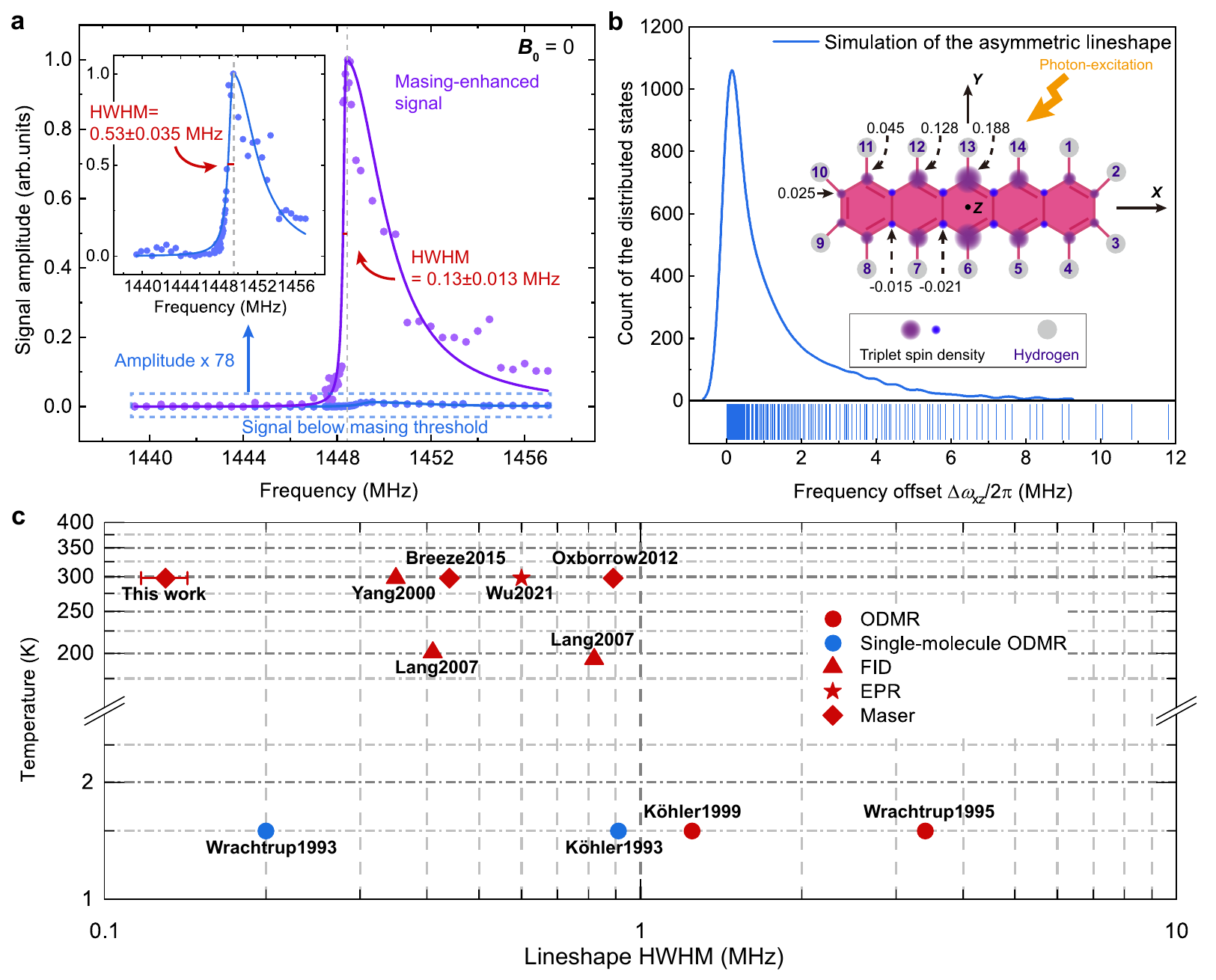}% Here is how to import EPS art
\caption{\label{fig: Lineshape_narrowing_211103}\textbf{Masing effects on the lineshape of pentacene's $\ket{X}\leftrightarrow\ket{Z}$ transition at zero field.} \textbf{a,} Comparison of the lineshapes obtained below (blue) and above (purple) the masing threshold. The measured signal amplitudes (dots) are plotted as a function of frequency and fitted with a bi-Lorentzian function (solid lines). The fitted half-width at half-maximum (HWHM) of the low-frequency component of each lineshape is labelled in red. The dashed grey lines indicate the peak positions. Inset: Zoom-in view of the below-threshold lineshape with a 78-fold magnification in its amplitude. \textbf{b,} Simulated asymmetric lineshape of pentacene's $\ket{X}\leftrightarrow\ket{Z}$ transition at zero field. Frequency offsets of the $2^{14}$ substates arising from the second-order hyperfine interaction are calculated and plotted in a kernel density curve with rug. The number of bins is 1000. Inset: The molecular structure of pentacene with the 14 protons (grey circles) labelled. The values of triplet spin densities (purple and blue circles) at different carbon nucleus are partially labelled due to the molecular symmetry. \textbf{c,} Comparison of the low-frequency-side HWHM of pentacene's zero-field $\ket{X}\leftrightarrow\ket{Z}$ transition lineshape measured in different studies. All low-temperature results\cite{kohler1993magnetic,wrachtrup1993optical,wrachtrup1995variation,kohler1999magnetic} are obtained using optically detected magnetic resonance (ODMR). The free-induction decay (FID)\cite{yang2000zero,lang2007dynamics}, electron paramagnetic resonance (EPR)\cite{wu2021bench} and masers\cite{oxborrow2012room,breeze2015enhanced} are employed to measure the HWHM at the higher temperatures ($\geq$ 194 K). Blue: single molecular spin; Red: spin ensemble. }
\end{figure*}

Firstly, we validate the above concept by simulating the effects of maser action on the magnetic resonance feature of a solid-state spin ensemble. The Lotka-Volterra model (see Methods) is employed to simulate the correlation between $N$ and $n$ in the presence of a high-Purcell-factor microwave resonator\cite{breeze2015enhanced}. The threshold behaviour is observed in Fig.~\ref{fig: Concept_Figure_211117}c which clearly reveals the below- and above-threshold regimes where two specific values of the number of inverted spins, $N=2\times10^{11}$ and $6\times10^{12}$ are chosen respectively for comparison. The slope at $N=6\times10^{12}$ is two orders of magnitude larger than that at $N=2\times10^{11}$ indicating the non-linearity of the correlation between $N$ and $n$. Therefore, above the threshold, a slight increase in inverted spins will result in an exponential increase in detected microwave photons. 

We apply the obtained threshold-like correlation to an inhomogeneously broadened spin ensemble with an intrinsic transition linewidth of 4 MHz. When all spin packets possess less inverted spins than $N_\textrm{threshold}$, there is no masing process during readout. This leads to the measured magnetic resonance lineshape, constituted by a distribution of detected microwave photons across different frequencies $n(\omega)$, identical to the intrinsic transition lineshape, as shown in the top panel of Fig.~\ref{fig: Concept_Figure_211117}d. In contrast to the below-threshold condition, for the spin ensemble with a central spin packet where $N=6\times10^{12}$ surpasses the masing threshold, the observed resonance linewidth is markedly reduced to 1 MHz (see bottom panel of Fig.~\ref{fig: Concept_Figure_211117}d) since the lineshape is dominated by the maser transitions occurring across a narrower range of frequencies. In addition to the linewidth narrowing, a 30-fold increase of the number of inverted spins, leading to a transition from the below- to the above-threshold regime, results in a boost of the number of microwave photons in the resonator by $\sim$450 times. Hence, the obtained ratio of the peak values of the microwave photons to inverted spins, $n_\textrm{max}/N_\textrm{max}$ increases by a factor of 15, which implies a non-linear enhancement of the magnetic resonance amplitude by masing.

The simulation results drove us to investigate the effects of maser action experimentally. Pentacene-doped \textit{p}-terphenyl is a suitable testbed due to its feasibility of masing at room temperature in solid states at zero field\cite{oxborrow2012room}. The maser action is achieved by exploiting the interaction between microwave photons and the photo-excited triplet spins of pentacene which are generated through the triplet mechanism demonstrated in Fig.~\ref{fig: Spin_system_setup}b. Upon photo-excitation, the introduced triplets of pentacene are highly polarized with a majority of populations in the upper sublevel $\ket{X}$ of the non-degenerate lowest triplet state T$_1$. Thus, the non-degeneracy arising from the zero-field splitting leads to an inverted two-level system consisting of the $\ket{X}$ and $\ket{Z}$ sublevels which permits a maser transition at $\sim$1.45 GHz at zero field. It is worth noting that, in contrast to the well-known inorganic solid-state spin ensembles comprising vacancy-type defect spins such as nitrogen vacancies (N$V$) in diamond\cite{breeze2018continuous} and silicon vacancies ($V_\textrm{Si}$) in silicon carbide (SiC)\cite{kraus2014room}, pentacene molecules are substitutionally doped in the lattice of \textit{p}-terphenyl (see Fig.~\ref{fig: Spin_system_setup}a) which allows relatively high doping concentrations of pentacene up to a few parts per thousand, thus favourably providing sufficient triplet spins of pentacene participating in the masing process. 

Herein, a pentacene-doped \textit{p}-terphenyl crystal with a doping concentration of 0.1\%, i.e. 1000 ppm, is prepared (see Methods) for the experiments. A frequency-tunable regenerative microwave oscillator, configured by a strontium titanate (STO) microwave resonator and a feedback loop (shown in Fig.~\ref{fig: Spin_system_setup}c), is employed as a zero-field magnetic resonance spectrometer for determining the resonance features of the sample under the different conditions discussed in the following context.

The zero-field magnetic resonance features of pentacene's $\ket{X}\leftrightarrow\ket{Z}$ transition are obtained below and above the masing threshold using different optical pump energies (see Methods). As demonstrated in Fig.~\ref{fig: Lineshape_narrowing_211103}a, the linearly normalized lineshape measured with masing reveals a prominent enhancement of the resonance amplitude by a factor of 78, in contrast to that detected below the threshold. As the number of pentacene's triplet spins is only increased by 50 times from the below- to above-threshold regime, this also confirms the non-linearity of the threshold behaviour and reveals the potential of maser action for non-linear enhancement of quantum-sensing sensitivity.

On the other hand, the transition lineshape is found to be asymmetrically distributed. The asymmetry is a characteristic feature of the zero-field transition of the triplet spins of pentacene doped in \textit{p}-terphenyl\cite{kohler1999magnetic}, which arises from the second-order hyperfine interactions ($2^\textrm{nd}$-order HFI) between the photo-excited triplet electron spins and the 14 protons of pentacene shown in Fig.~\ref{fig: Lineshape_narrowing_211103}b. The asymmetry of the $\ket{X}\leftrightarrow\ket{Z}$ transition lineshape can be theoretically predicted by a second-order perturbation formalism\cite{kohler1999magnetic} (see Methods for details) which qualitatively provides the distributions of the $2^\textrm{nd}$-order-HFI-modified energy states of $\ket{X}$ and $\ket{Z}$.   

Besides the signal enhancement, the other predicted effect of masing-mediated linewidth narrowing on the magnetic resonance feature is also confirmed. Due to the asymmetry of the measured resonance lineshapes shown in Fig.~\ref{fig: Lineshape_narrowing_211103}a, a bi-Lorentzian fitting is employed to determine the linewidth. We find that the full-width at half-maximum (FWHM) of the pentacene's $\ket{X}\leftrightarrow\ket{Z}$ transition lineshape is reduced from $3.3\pm0.48$ MHz to $2\pm0.17$ MHz by maser action. More surprisingly, the steep low-frequency side of the asymmetric lineshape with a half-width at half-maximum (HWHM) of $530\pm35$ kHz is narrowed by 4 fold to $130\pm13$ kHz. To the best of our knowledge, such a narrow HWHM is rarely achieved in solid-state spin ensembles. Comparing with the reported HWHM values of the same spin transition obtained by different techniques\cite{wrachtrup1993optical,kohler1993magnetic,wrachtrup1995variation,kohler1999magnetic,yang2000zero,lang2007dynamics,oxborrow2012room,breeze2015enhanced,wu2021bench}, the HWHM obtained herein with the aid of masing is the narrowest (see Fig.~\ref{fig: Lineshape_narrowing_211103}c). Notably, it is 60\% of the measured HWHM of a \textit{single} pentacene molecule in \textit{p}-terphenyl at \textit{cryogenic} temperatures\cite{wrachtrup1993optical}. Despite two previous works\cite{oxborrow2012room,breeze2015enhanced} studying the same zero-field transition in the presence of masing as well, the reported HWHM values are slightly larger as they are dominated by the relatively broad linewidth of their resonators, whereas in our case, instead of using a microwave resonator, we configure a microwave oscillator with an extremely narrow linewidth which offers more accurate frequency selection and better spectral resolution for measuring the magnetic resonance features.

\begin{figure*}[htbp!]
\includegraphics{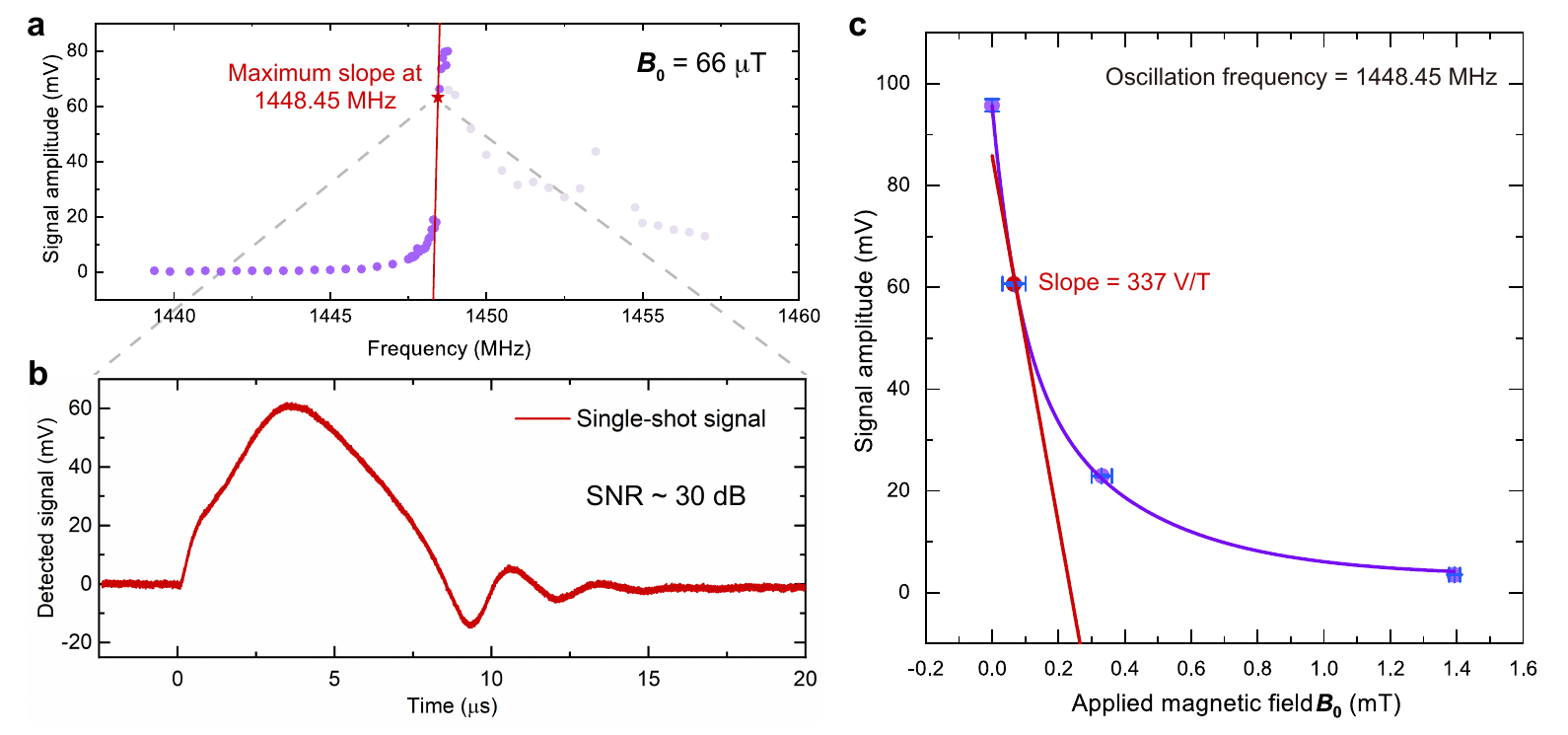}% Here is how to import EPS art
\caption{\label{fig: Magnetic_field_sensing_211118}\textbf{The performance of magnetic-field sensing.} \textbf{a,} Signals of pentacene's $\ket{X}\leftrightarrow\ket{Z}$ transition measured with above-threshold pumping at \textbf{\textit{B}$_\textrm{0}$} = 66 $\mu\textrm{T}$. The signals are plotted as a function of frequency (purple dots), in which the steepest rise in signal amplitude, i.e. the maximum slope is determined at 1448.45 MHz (red star). \textbf{b,} A single-shot time-domain signal obtained at 1448.45 MHz. The signal-to-noise ratio (SNR) is labelled. \textbf{c,} Magnetic-field dependence of the signal amplitude measured at 1448.45 MHz. The data are fitted with an exponential curve (purple) to which a tangent line is employed to estimate the slope at \textbf{\textit{B}$_\textrm{0}$} = 66 $\mu\textrm{T}$. The vertical error bars are obtained by fitting the peaks of the time-domain signals. The horizontal error bars are the s.d. of the calibrated \textbf{\textit{B}$_\textrm{0}$}.  }
\end{figure*}

Since the observed masing-enhanced magnetic resonance features are advantageous to quantum sensing with the slope detection protocol\cite{degen2017quantum} e.g. magnetometry\cite{barry2020sensitivity}, we study the feasibility of the spin ensemble (i.e. pentacene-doped \textit{p}-terphenyl) acting as an organic solid-state magnetometer operating at ambient conditions. The working principle of our magnetometer is similar to the continuous-wave optically detected magnetic resonance (cw-ODMR) magnetometry\cite{barry2020sensitivity}, but differs in the type of readout signals and optical initialization. Here, we detect microwave signals (boosted by masing) instead of optical ones and our optical initialization, i.e. optical pumping for generating the pentacene's triplet spins that needs only several nanoseconds, is at least one order of magnitude faster than that required in optical readout schemes. 

Using the same microwave circuitry demonstrated in Fig.~\ref{fig: Spin_system_setup}c, we find that in the presence of a dc bias magnetic field $\textbf{\textit{B}$_\textrm{0}$}$ of 66 $\mu$T, the magnetic resonance feature (see Fig.~\ref{fig: Magnetic_field_sensing_211118}a) shows a maximum slope at 1448.45 MHz where the resonance signal is most sensitive to changes in magnetic field. Owing to the masing process, the single-shot signal read out at 1448.45 MHz has a signal-to-noise ratio of 30 dB as shown in Fig.~\ref{fig: Magnetic_field_sensing_211118}b. The Rabi oscillations are attributed to the coherent energy exchange between the spin ensemble and numerous microwave photons stored in the microwave oscillator which are normally observed in STO-based high-power pentacene masers\cite{breeze2017room,wu2021bench}. 

By monitoring changes of the readout signal at 1448.45 MHz under different applied magnetic fields, we evaluate the sensitivity of the pentacene-based magnetometer. Fig.~\ref{fig: Magnetic_field_sensing_211118}c shows incremental changes of magnetic field around 66 $\mu$T resulting in maximal changes in the signal amplitude and the correlation is represented by a slope $m_\textrm{s}$ equal to 337 V/T. By convention, the sensitivity $\eta$, defined as the readout signal with a SNR of 1\cite{taylor2008high,barry2020sensitivity}, can thus be estimated according to\cite{schloss2018simultaneous}
\begin{equation}\label{sensitivity}
    \eta = \frac{\sigma_\textrm{s}}{m_\textrm{s}\sqrt{2\Delta f}}
\end{equation}
where $\sigma_\textrm{s}=4.5\times10^{-4}$ V is the single-shot standard deviation of the off-resonance time-domain signal measured at 66 $\mu$T and $\Delta f=500$ MHz is the measurement bandwidth used for our data acquisition. Therefore, the sensitivity of the pentacene-based magnetometer is projected to be 42 pT/$\sqrt{\textrm{Hz}}$ which is comparable to that of the state-of-the-art solid-state ensemble magnetometers such as N$V$-based magnetometers\cite{barry2020sensitivity}. We here propose the following immediate strategies which can be implemented to improve the sensitivity. First, the bias dc magnetic field can be increased to enlarge the energy splittings of the pentacene's triplet sublevels which will improve the sensor by providing a linear zone in which the sensitivity is not limited by the zero-field splittings. Although the resonance linewidth will be broadened by the increased magnetic field\cite{xie2018mesoscopic}, the pentacene can be deuterated to suppress the hyperfine interactions, so compensating for the broadening\cite{kohler1999magnetic}. Moreover, careful tuning of the optical pump energy may give rise to an optimal condition that only triplet spin packets in the extreme vicinity of the resonance reach the masing threshold, leading to further reduction of the magnetic resonance linewidth.

Our study validates the concept of exploiting maser action in solid-state ensemble spin sensors for improvement of the sensitivity. This technique offers a promising path to mitigate the downsides of inhomogeneous broadening inherent in the sensitivity of solid-state ensemble spin sensors. The masing-enhanced readout is advantageous over conventional optical readout in terms of the single-shot SNR and overhead time. We note that our readout protocol would seem to enrich the rising field of microwave readout\cite{eisenach2021cavity,ebel2021dispersive} for quantum sensing and enables microwave readout with (i) a higher SNR; (ii) an appealing threshold-like (non-linear) enhancement of the sensitivity $\eta$ with increasing number of spins $N$, whereas the dependence of $\eta$ on $N$ is linear and square root for alternative microwave readout\cite{ebel2021dispersive,eisenach2021cavity} and conventional optical readout protocols, respectively. Furthermore, this technique is fully compatible with the extensively studied solid-state spin sensors based on N$V$ diamond and $V_\textrm{Si}$ in SiC, which are both candidates for room-temperature solid-state masers\cite{jin2015proposal,breeze2018continuous,kraus2014room}, and it can also be used to explore exotic quantum sensors (e.g. the organic solid-state magnetometer demonstrated herein) for broad applications in electric/magnetic field sensing\cite{degen2017quantum} and dark matter searches\cite{jiang2021search}.
\\
\begin{acknowledgments}
\noindent We thank Wern Ng, Min Jiang, Tianyu Xie for valuable discussions. This study was supported by NSF of China (Grant
No. 12004037, No. 91859121), the National Key R\&D Program of China (Grant No. 2018YFA0306600), Beijing Institute of Technology Research
Fund Program for Young Scholars, U.K. Engineering and Physical Sciences Research Council (Grant No. EP/K037390/1, No. EP/M020398/1) and the China Postdoctoral Science Foundation (Grant No. YJ20210035, No. 2021M700439).
\end{acknowledgments}

\newpage
\appendix

\section{Methods}

\noindent \textbf{Simulation of maser-mediated linewidth narrowing and signal enhancement.} The maser-mediated linewidth narrowing and signal enhancement are simulated based on the Lotka-Volterra model\cite{lotka1920undamped}, which comprises a pair of first-order non-linear differential equations, as shown below:
\begin{equation}\label{LVmodel}
\begin{split}
    &\Dot{N} = -2BnN-\gamma_\textrm{s} N \\
    &\Dot{n} = -\kappa_\textrm{c}n+BnN
\end{split}
\end{equation}
where $N$ and $n$ are the number of the inverted spins and microwave photons in a resonator's electromagnetic mode, and their decay rates are expressed by $\gamma_\textrm{s}$ and $\kappa_\textrm{c}$, respectively. $B$ is the Einstein coefficient of the stimulated emission. The values of these parameters are adopted from previous studies on the pentacene's triplet spin system, where $\gamma_\textrm{s} = 4\times10^4$ s$^{-1}$ taking both processes of depopulation and spin-lattice relaxation into account\cite{wu2019unraveling}, $B=11\times10^{-8}$ s$^{-1}$ and $\kappa_\textrm{c}=2.1\times10^6$ s$^{-1}$ are typical parameters of strontium titanate (STO)-based pentacene masers\cite{salvadori2017nanosecond,wu2020room}. For simplicity, the model is solved in steady state, i.e. $\Dot{N}=\Dot{n}=0$. Thus, under the conditions of $n$ and $N>0$, we arrive at the threshold-like correlation (as depicted in Fig.~\ref{fig: Concept_Figure_211117}c) between the number of inverted spins and microwave photons:
\begin{equation}\label{nNcorrelation}
    n = \frac{\gamma_\textrm{s}N}{\kappa_\textrm{c}-3BN}
\end{equation}

Based on equation~(\ref{nNcorrelation}), we simulate the masing effect on an inhomogeneously broadened spin ensemble with a Lorentzian lineshape and intrinsic linewidth of 4 MHz. The lineshape can be represented by a distribution of $N$ through different frequencies, i.e. $N(\omega)$. The below- and above-threshold regimes are achieved by setting the peak value of the lineshape $N_\textrm{max}=2\times10^{11}$ and $6\times10^{12}$, respectively. Combining with equation~(\ref{nNcorrelation}), the distribution of $n(\omega)$ is simulated in both regimes revealing the masing effect on the observed linewidth and signal amplitude. 
\\

\noindent \textbf{Preparation of pentacene-doped \textit{p}-terphenyl crystals.} The Bridgman method\cite{bridgman2013certain} is employed to grow pentacene-doped \textit{p}-terphenyl crystals. Raw \textit{p}-terphenyl (Alfa Aesar, 99+\%) is purified by a home-built zone refiner set at 230 $^\circ$C. The process of zone refining takes 30 iterations, lasting 4 days. Pentacene (TCI Chemicals, purified by sublimation) and purified \textit{p}-terphenyl powders are mixed in a molar ratio of 1 to 1000 and sealed into a borosilicate tubing under argon. The tubing is loaded inside a vertical furnace (Elite Thermal Systems Ltd.) in which the mixed powders are melted at 217 $^\circ$C. The melt is then lowered to a cold zone for crystallization with an optimized rate of 2 mm/h. Once the crystallization is completed, the crystal grown is slowly cooled to room temperature with a rate of 10$^\circ$C/h to avoid thermal shock. Removed from the tubing, a 0.1\% pentacene-doped \textit{p}-terphenyl crystal (height 7.2 mm, width 4 mm and average thickness 1.5 mm) is cut for the experiments. 
\\

\noindent \textbf{Design of microwave dielectric resonators.} The microwave dielectric resonator comprises a cylindrical oxygen-free copper cavity (inner diameter 40 mm, internal height 31 mm) and a dielectric ring made of monocrystalline STO (inner diameter 5 mm, outer diameter 12 mm, height 7.5 mm, all surfaces finely polished). The STO ring, placed concentrically in the cavity, is lifted by 10.5 mm above the floor of the cavity via a Rexolite support. A "piston-like" copper tuning screw, at the top of the cavity, is used to change the distance between the STO ring and the cavity's ceiling so as to adjust the TE$_{01\delta}$ mode's frequency in the resonator. A 4-mm hole is drilled in the cavity's wall for firing a laser beam into the cavity to pump the pentacene-doped \textit{p}-terphenyl crystal optically. Two loop antennas are inserted inside the cavity functioning as an under-coupled input port and an over-coupled output port, respectively. The loaded ($Q_\textrm{L}$) and unloaded ($Q_0$) quality factors of the microwave dielectric resonator are determined via a microwave analyzer (Keysight N9917A) to be $Q_\textrm{L}\approx800$ and $Q_\textrm0 \approx 7000$.
\\

\noindent \textbf{Lineshape measurements.} The prepared 0.1\% pentacene-doped \textit{p}-terphenyl crystal is loaded inside the bore of the STO ring with its cleavage plane vertically aligned. The microwave resonator connected with a feedback loop acts as a regenerative microwave oscillator as demonstrated in Fig.~\ref{fig: Spin_system_setup}c. The power and frequency of self-oscillation in the microwave circuitry are regulated by the variable attenuator, phase shifter and tuning screw of the resonator, collectively. For lineshape measurements, the microwave power inputted into the resonator is set to be -9 dBm and the frequency is varied from 1439.375 to 1457 MHz, which are monitored by the microwave analyzer (Keysight N9917A). An optical parametric oscillator (OPO) (Deyang Tech. Inc. BB-OPO-Vis, pulse duration 7 ns) pumped by a Nd:YAG $Q$-switched laser (Beamtech Nimma-900, repetition rate 10 Hz) is used for optically pumping the gain medium (i.e. pentacene-doped \textit{p}-terphenyl). The OPO output (wavelength 590 nm) is focused to 2 mm and enters the cavity through the 4-mm hole in its side wall. The data in Fig.~\ref{fig: Lineshape_narrowing_211103}a and \ref{fig: Magnetic_field_sensing_211118}a are obtained with the optical pulse energy of 4 mJ, except for the below-threshold measurements (inset of Fig.~\ref{fig: Lineshape_narrowing_211103}a) in which 80-$\mu$J pulses are used. 

Upon photo-excitation of the gain medium, we sweep the frequency of the oscillator in steps of 500 kHz when it is far from the resonance and steps of 50 kHz when we approach the resonance. The signals reflected as power changes of the oscillator are measured by a logarithmic detector (AD8317, scale 22 mV/dB) AC-coupled to a digital storage oscilloscope (Tektronix MSO64, sampling rate 6.25 GSa/s). The laser pulses monitored by a photodetector (Thorlabs, DET10A2) (not shown in Fig.~\ref{fig: Spin_system_setup}c) are employed to trigger the oscilloscope. Due to the limited dynamic range (55 dB) of the logarithmic detector and dramatic signal enhancement by the masing process, 10 or 35-dB attenuation is added at the input of the detector to avoid signal saturation and/or damage on the detector. All signals near resonance (i.e. masing signals) are collected by single-shot measurements while weak far-off-resonance signals are averaged and then collected. The background signals introduced by laser-induced microphonics are collected and subtracted when a $\sim$100 G static magnetic field is applied on the crystal by a neodynium-iron-boron permanent magnet, which completely shifts pentacene's $\ket{X}\leftrightarrow\ket{Z}$ transition frequency beyond the resonator's bandwidth. The peak values of the time-domain signals are plotted as a function of frequency which manifest the lineshapes. The lineshapes are converted from a logarithm scale to a linear scale and normalized for obtaining the HWHM with a bi-Lorentzian fitting.
\\

\noindent \textbf{Simulation of zero-field asymmetric lineshape.} The asymmetric lineshape of pentacene's $\ket{X}\leftrightarrow\ket{Z}$ transition can be qualitatively interpreted by calculating the energy shifts of the $\ket{X}$ and $\ket{Z}$ spin levels arising from the second-order hyperfine interactions (2$^\textrm{nd}$-order HFI). Due to the smallest zero-field splitting (ZFS) between the $\ket{X}$ and $\ket{Y}$ spin levels, the second-order energy shift of $\ket{X}$ is much more pronounced than that of $\ket{Z}$. Thus, with energy in units of MHz, the shift of $\ket{X}\leftrightarrow\ket{Z}$ transition frequency can be estimated:
\begin{equation}\label{frequencyshift}
\Delta\omega_\textrm{XZ}/2\pi = \Delta E_\textrm{X}-\Delta E_\textrm{Z} \approx \Delta E_\textrm{X}
\end{equation}
According to ref.~\onlinecite{kohler1999magnetic}, one obtains
\begin{equation}\label{asymmetric}
\begin{split}
    &\Delta E_\textrm{X} = \frac{1}{4}\frac{|\pm\rho_1A_\textrm{ZZ}^{(1)}\pm\rho_2A_\textrm{ZZ}^{(2)}\pm\cdot\cdot\cdot\pm\rho_{14}A_\textrm{ZZ}^{(14)}|^2}{E_\textrm{X}-E_\textrm{Y}}\\
\end{split}
\end{equation}
where $\rho_\textrm{i}$ ($\textrm{i}=1,2,\cdot\cdot\cdot,14$) is the triplet spin density\cite{keijzers1989pulsed} at the carbon nucleus bounded with the 14 protons (labelled in Fig.~\ref{fig: Lineshape_narrowing_211103}(b)), $A_\textrm{ZZ}^{(\textrm{i})}=-61$ MHz\cite{atherton1993principles} is the associated proton-hyperfine tensor elements where both isotropic and anisotropic parts are included, $E_\textrm{X}-E_\textrm{Y}=107.5$ MHz\cite{yang2000zero} is the ZFS between the $\ket{X}$ and $\ket{Y}$ spin levels and the "$\pm$" sign implies the two possible states of the protons. The asymmetric lineshape of pentacene's $\ket{X}\leftrightarrow\ket{Z}$ transition at zero field is thus simulated by plotting the $2^{14}$ solutions of equation~(\ref{asymmetric}) in a kernel density curve with rug, as shown in Fig.~\ref{fig: Lineshape_narrowing_211103}(b).
\\

\noindent \textbf{Sensing of external magnetic fields.} We perform the lineshape measurements while a small dc magnetic field is applied. The magnetic field is provided by a neodynium-iron-boron permanent magnet and varied by changing the distance between the gain medium and magnet. The applied magnetic field is calibrated by a commercial magnetometer (CH-1600, measurement precision 10 nT). Its transverse probe (active area < 1 mm$\times$2.6 mm) is inserted inside the bore of the STO ring and moved along the ring's cylindrical axis in steps of 0.1 mm to map the homogeneity of the magnetic flux perpendicular to the gain medium's cleavage plane.

\nocite{*}


\providecommand{\noopsort}[1]{}\providecommand{\singleletter}[1]{#1}%

\end{document}